%% file: paper.tex
\documentclass[
superscriptaddress,
nofootinbib,
 amsmath,
 amssymb,
 aps,
prb,
floatfix,
]{revtex4-2}

\usepackage{graphicx}% Include figure files
\usepackage{dcolumn}% Align table columns on decimal point
\usepackage{bm}% bold math
\usepackage[utf8]{inputenc}
\usepackage{tabularx}
\usepackage[dvipsnames]{xcolor}
\usepackage[euler]{textgreek}
\usepackage{upgreek}
\usepackage{derivative}
\usepackage{amsmath}
\usepackage{physics}
\usepackage{amssymb}
\usepackage{hyperref}
\hypersetup{colorlinks=true,linkcolor=blue!50!black,urlcolor=blue!50!black,citecolor=blue!50!black}
\usepackage{colortbl}
\usepackage{booktabs}
\usepackage{array}
\usepackage{makecell}
\usepackage{subfigure}
\usepackage[margin=1in]{geometry}  % To control margins
\usepackage{dsfont}
\usepackage{feynmp-auto}
\usepackage{float}
\usepackage{color}
\usepackage{physics}

\usepackage{tikz}
\usepackage{tikz-feynman}
\tikzfeynmanset{compat=1.1.0} % or latest installed version

\newcommand\freefootnote[1]{%
  \let\thefootnote\relax%
  \footnotetext{#1}%
  \let\thefootnote\svthefootnote%
}
\usepackage{siunitx}

\begin{document}
\preprint{APS}

\title{Cosmic Ray Boosted Dark Matter in COSINUS: Modeling and Constraints}

\input{COSINUS.tex}

\newcommand{\Tx}{T_\mathrm{\chi}}
\newcommand{\Ti}{T_i}
\newcommand{\mx}{m_\mathrm{\chi}}
\newcommand{\mi}{m_i}
\newcommand{\Er}{E_\mathrm{R}}

\begin{abstract}
Direct detection of nuclear recoils due to sub-GeV dark matter is challenging because of the small kinetic energy of the light dark matter particles. Although limits down to a few hundred MeV have been reached using specially designed low threshold detectors, further improvements are now constrained more by background event rates than by energy thresholds. However, constraints down to sub-MeV dark matter masses can still be obtained through the boosted dark matter framework. In this scenario, high-energy cosmic rays or neutrinos scatter off dark matter particles, imparting additional kinetic energy and boosting them beyond the typical velocities expected from the non-relativistic dark matter halo. These boosted dark matter particles can then be detected even by experiments with higher energy thresholds. In this work, we present a catalog of dark matter - nucleon scattering cross sections corresponding to a heavy mediator limit for spin $0, \frac{1}{2}$ and 1 dark matter and for scalar and vector mediators with even or odd parity. Based on these results, we present projected constraints on the dark matter–nucleon cross section for the COSINUS experiment, assuming an exposure of \SI{100}{kg\,d}, demonstrating the potential sensitivity to sub-GeV boosted dark matter.
\end{abstract}

\maketitle
%%%%%%%%%%%%%%%%%%%%%%%%%%%%%%%%%%%%%%%%%%%%%%%
%%%%%%%%%%%%%%%%%%%%%%%%%%%%%%%%%%%%%%%%%%%%%%%
\section{Introduction}
\label{introduction}
The search for non-luminous gravitationally interacting matter, dark matter (DM), has been the focus of many experimental setups in the past decades. Cosmological observations on different length scales hint at the existence of such new particles. Anomalies in rotation curves (kpc) \cite{sofue_rotation_2001}, gravitational lensing or galaxy cluster redshifts (Mpc) \cite{clowe_direct_2006, harvey_non-gravitational_2015} and the CMB (Gpc) \cite{calabrese_atacama_2025, collaboration_planck_2020} all demonstrate a presence of invisible mass within the universe. There has been a wide effort to detect the DM particles in laboratory environments, but so far the fundamental properties of DM remain a mystery.
Direct detection experiments have reported no significant DM induced signals until this day \cite{Schumann:2019eaa,Billard:2021uyg,Misiaszek:2023sxe}. The most stringent limit is from the LUX-ZEPLIN collaboration who, using Xenon as target material, were able to constrain the DM-nucleon cross section down to \SI{2.1e-48}{cm^2} for a DM particle of mass \SI{36}{GeV} \cite{aalbersDarkMatterSearch2025}. 
Based on theory of cosmic structure formation and on astrophysical observations, the DM in the Milky Way halo is assumed to be non-relativistic, with typical velocities\footnote{Throughout this paper we work in natural units, $c=1$.} of the order of $v\sim 10^{-3}$. Therefore, DM masses above several GeV are needed in order to induce a nuclear recoil above the typical few keV threshold of the detectors. 
Hence, most experiments, not specifically designed for low threshold measurements, drastically lose sensitivity in the lower-mass regions, below a few GeV. The loss of sensitivity is determined by the energy threshold of the experiment; a lower threshold provides stronger constraints in the region of smaller DM masses. 
However, if the DM particles can scatter off nuclei in the direct detection experiment, scattering processes between DM and cosmic ray (CR) particles must also be possible. Given the energetic nature of the CR particles, these processes will boost the DM particles, resulting in increased DM kinetic energies. Thus, boosted DM can overcome the constraints set by high thresholds, compensating the sub-GeV nature of the DM with appropriate high energies \cite{bringmannNovelDirectDetection2019}. 

Previously, ideas about boosted dark matter (BDM) have been explored e.g., by COSINE \cite{COSINE-100:2023tcq}, LZ \cite{LZ:2025iaw} and PANDA-4X \cite{PandaX:2024pme} setting constrains on sub-GeV DM. The boost obtained by the DM particles can be received through various mechanisms. COSINE \cite{COSINE-100:2023tcq} conducted a search assuming annihilation processes resulting in two boosted light DM particles. LZ \cite{LZ:2025iaw} and PANDA-4X \cite{PandaX:2024pme} assume elastic scattering between the CR particles and the DM particles of the halo. However, the former assumes higher energetic extragalactic CR particles, while the latter assumes galactic electrons. In this work, we will use the flux obtained from extragalactic CR particles, following previous literature \cite{LZ:2025iaw, Ghosh:2024dqw}.
In addition to the nuclear interactions, the DM particles might be boosted by neutrinos. However, this requires interactions beyond the DM-nucleon coupling which is the basis of the nuclear recoil in direct detection experiments. Assuming DM-neutrino cross sections can improve the bounds obtained through BDM \cite{Ghosh:2024dqw, Ghosh:2021vkt}. Hence, in this work we will explore the impact of such interactions and draw theoretical bounds. However, we will not consider DM-electron interactions, as a detailed analysis of electron scattering sensitivity in the COSINUS experiment is yet to be completed. 

The COSINUS experiment features Sodium Iodine (NaI) cryogenic scintillating calorimeters located at the Laboratori Nazionali del Gran Sasso (LNGS) deep underground facilities in Italy. The detectors are operated at low temperature (<$\SI{10}{mK}$) allowing to read phonon signals emitted by the crystal lattice in combination with scintillation light \cite{angloherCOSINUSModelindependentChallenge2025, angloherFirstMeasurementsRemoTES2023}. The dual readout of phonon and light signals enables event-by-event particle discrimination, allowing to distinguish between electron and nuclear recoil events. The background contributions are minimized by multiple shielding procedures installed \cite{Angloher:2024pbw}, hence in combination with the particle discrimination enabling an almost background free DM-nucleon scattering environment. For the initial run, the intended exposure, given by detector mass multiplied by run duration, is $\SI{100}{kg\,d}$. The target nuclear recoil energy threshold is $\SI{1}{keV}$ and the baseline energy resolution $\SI{0.2}{keV}$\cite{angloherCOSINUSModelindependentChallenge2025}.

In this work, we investigate the constraints on the DM-nucleon cross section for sub-GeV DM in the context of BDM in the COSINUS experiment. First, we will apply a model that has been commonly referenced in BDM studies, corresponding to an energy independent contact interaction between the DM and the nucleon. 
Then, we will consider simplified models that systematically categorize the DM-nucleon interactions in the heavy mediator limit; we explore the scenarios where the DM particle is a scalar, a fermion or a vector boson, and the mediator is a scalar, pseudoscalar, vector or an axial vector boson. We observe that all of the heavy mediator scenarios, with the exception of the pseudoscalar mediator case, lead to stronger constraints compared to the energy-independent contact interaction limit. Hence, the contact interaction scenario can be used as a conservative, almost model-independent estimate.

\section{Boosted Dark matter flux}
The velocity dispersion in the Standard Halo Model (SHM) is of the order of $v \sim 10^{-3}$, implying that DM particles are fundamentally non-relativistic. Compared to the velocities of CR particles, which can be highly relativistic, the DM particle is approximately at rest in the galactic frame. We label the kinetic energies  $T_i$ for the initial state CR particle and $T_\mathrm{\chi}$ for the final state BDM particle, and take the kinetic energy of the initial state DM particle to be zero. The subscript $i$ denotes the particle type, namely the proton and Helium $i \in \{\mathrm{p}, ^4$He$\}$. The maximal kinetic energy of the DM particle after the scattering process is given by
\begin{align}
    \label{eq:TxMax}
    \Tx^\mathrm{max} = \frac{T_i^2 + 2 \mi \Ti}{\Ti + (\mi + \mx)^2 / (2 \mx)}.
\end{align}
Here, $\mx$ ($\mi$) denotes the DM (CR) particle's mass. Consequently, the minimum required kinetic energy of the CR particle to boost a DM particle to energy $\Tx$ is obtained by inverting \eqref{eq:TxMax}, 
\begin{align}
    \label{eq:TiMin}
    \Ti^\mathrm{min} = \left(\frac\Tx2 - \mi\right)\left[1 \pm \sqrt{1 + \frac{2 \Tx (\mi + \mx)^2}{\mx (2 \mi - \Tx)^2}}\right].
\end{align}
The sign is determined by the relation between the kinetic energy and the CR particle's mass, namely for $+$ ($-$) we have the condition $\Tx > 2 \mi$ ($\Tx < 2 \mi$) \cite{bringmannNovelDirectDetection2019, Ghosh:2024dqw, bellCosmicrayDarkMatter2024}. 

\begin{figure}[H]
    \centering
    \includegraphics[width=.7\linewidth]{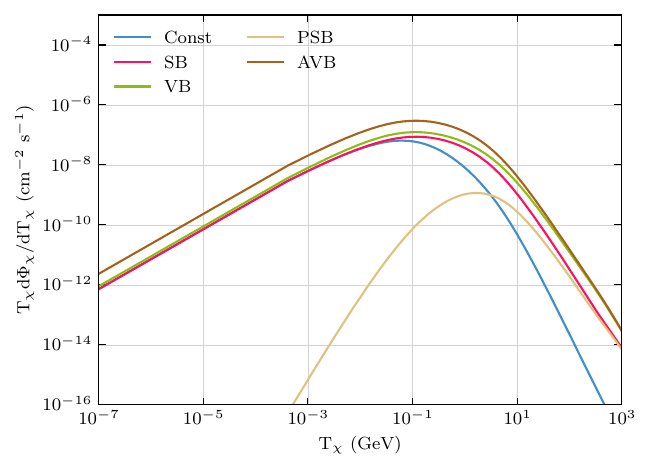}
    \caption{BDM flux for fermion DM considering the different heavy mediator cases as well as the energy independent interaction (blue). The mediators include scalar (red), vector(green), pseudoscalar (beige) and axial vector (brown). The BDM flux are obtained from equations \eqref{eq:constCrossSection} - \eqref{eq:FMAVBCrossSection} and \eqref{eq:fluxBDM}. The reference cross section is chosen as $\bar\sigma_{\chi i} = 10^{-30}\,\mathrm{cm}^2$ and the DM mass $m_\chi = \SI{0.1}{GeV}$.}
    \label{fig:fluxBDMNucleons}
\end{figure}

The BDM flux can be obtained from the local interstellar spectra (LIS) \cite{DellaTorre:2016jjf, Boschini:2017fxq, Boschini:2020jty} and the DM-CR particle cross section.
The LIS describes the flux of the CR particles within and outside the heliosphere. The LIS is given as the differential flux with respect to the CR particle's energy. Note, the differential LIS flux is typically given per steradian and per rigidity, where we assume the flux to be isotropic. The latter can be transformed to kinetic CR energies as described in \cite{bringmannNovelDirectDetection2019}. Thus, we can obtain the differential CR flux with respect to the CR kinetic energy $d\Phi_\mathrm{\mathrm{LIS}}/d \Ti$. The differential BDM flux after being scattered by CR particles is given by
\begin{align}
    \frac{d\Phi_\mathrm{\chi}}{d \Tx} = D_\mathrm{eff} \frac{\rho_\chi}{\mx} \int_{\Ti^{\mathrm{min}}}^\infty \frac{d\Phi_\mathrm{\mathrm{LIS}}}{d \Ti} \frac{d \sigma_{\chi i}}{d \Tx} d\Ti,
\end{align}
where $D_\mathrm{eff} = \SI{8.02}{kpc}$ \cite{bringmannNovelDirectDetection2019} is the effective distance, $\rho_\chi = \SI{0.3}{GeV/cm^3}$ the local DM density and ${d \sigma_{\chi i}}/{d \Tx}$ the differential DM-nucleon cross section. The effective distance, also known as the diffusion zone, is obtained by assuming a NFW density profile and carrying out the line of sight integral up to $\SI{10}{kpc}$. The details of the underlying DM particle physics model are encoded in the differential cross section. Previously other works have examined the case of energy independent contact interaction cross section. Energy dependent cross sections have been investigated in the case of fermion (FM) DM interacting via a scalar boson (SB) or vector boson (VB) exchange with the CR particles \cite{Ghosh:2024dqw, bringmannNovelDirectDetection2019, maityCosmicrayBoostedDark2024}. Pseudoscalar (PSB) and axial vector (AVB) mediators have been analyzed in \cite{bellCosmicrayDarkMatter2024}. The energy independent differential cross section is given by
\begin{align} \label{eq:constCrossSection}
    \left(\frac{d \sigma_{\chi i}}{d \Tx}\right)_{\mathrm{const.}} = \frac{\bar\sigma_{\chi i}}{\Tx^\mathrm{max}},
\end{align}
where $\bar\sigma_{\chi i}$ is a reference DM-nucleon cross section, corresponding to the limit of zero momentum transfer. We can further obtain the energy dependent differential cross section for the FM case, as discussed in appendices \ref{app:cs-formalism} and \ref{Appendix:fermion}. Following this, consistent with \cite{Ghosh:2024dqw, bellCosmicrayDarkMatter2024}, we find
\begin{widetext}
    \begin{align}
    &\left(\frac{d \sigma_{\chi i}}{d \Tx}\right)_{\mathrm{FM}, \mathrm{SB}} = \frac{\bar\sigma_{\chi i} \mx}{4 s\Tx^\mathrm{max} \mu_{\chi i}^2}\left(2 \mx + \Tx\right) \left(2 \mi^2 + \mx \Tx\right), \label{eq:FMSBCrossSection}\\ %END    
    &\left(\frac{d \sigma_{\chi i}}{d \Tx}\right)_{\mathrm{FM}, \mathrm{PSB}} =\frac{\bar\sigma_{\chi i}\mx^2\Tx^2}{4 s\Tx^\mathrm{max} \mu_{\chi i}^2}, \label{eq:FMPSBCrossSection}\\ %END
    &\left(\frac{d \sigma_{\chi i}}{d \Tx}\right)_{\mathrm{FM}, \mathrm{VB}} = \frac{\bar\sigma_{\chi i} \mx}{2 s \Tx^\mathrm{max} \mu_{\chi i}^2} \Bigl[2 \mx \left(\mi + \Ti\right)^2 -  \Tx\left((\mi + \mx)^2 + 2 \mx \Ti\right)+ \mx \Tx^2\Bigr], \label{eq:FMVBCrossSection}  \\  %END
    &\left(\frac{d \sigma_{\chi i}}{d \Tx}\right)_{\mathrm{FM}, \mathrm{AVB}} = \frac{\bar\sigma_{\chi i} \mx}{2 s\Tx^\mathrm{max} \mu_{\chi i}^2} \Bigl[2 \mx \left(2 \mi \Ti + 3 \mi^2 \Ti^2 + \Ti^2\right) + \Tx \left(\left(\mi - \mx\right)^2 - 2 \mx \Ti\right) + \mx \Tx^2\Bigr], \label{eq:FMAVBCrossSection}
\end{align}
\end{widetext}
where $s$ is the center of mass energy squared and \mbox{$\mu_{\chi i}=m_{\chi}m_i/(m_\chi + m_i )$} the reduced mass of the DM and CR particle. The center of mass energy squared in the galactic frame is given as 
\begin{align}
    s = (\mx + \mi)^2 + 2 \mx \Ti.
    \label{eq:MandelstamS}
\end{align}
An underlying assumption in these equations is that the CR particles are pointlike. Although this is a valid assumption in the case of electron CR scattering processes, its not valid for baryons and nuclei. Hence, we need to consider a form factor. We use the dipole form factor as has been done in previous literature, given by
\begin{align}
\label{eq:formfactor}
    G_i^2(q^2) = \left(A_i\frac{\mu_{\chi i}}{\mu_{\chi n}}\right)^2\frac{1}{(1 + q^2/\Lambda_i^2)^2}.
\end{align}
The cutoff $\Lambda_i$ is given as $\Lambda_\mathrm{p}=\SI{0.77}{GeV}$ ($\Lambda_\mathrm{He}=\SI{0.41}{GeV}$) \cite{bringmannNovelDirectDetection2019, Ghosh:2024dqw, maityCosmicrayBoostedDark2024} for a CR proton (Helium). The momentum transfer squared is obtained in the galactic frame as $q^2 = 2 \mx \Tx$. The factor $A_i^2$ is the nucleon number squared of the element $i$, $\mu_{\chi i}$ the reduced DM CR particle mass and $\mu_{\chi n}$ the reduced DM nucleon mass. For protons, this factor simplifies to $A_p^2 \mu_{\chi p}/\mu_{\chi n} = 1$. Further, we assume equal couplings to protons and neutrons. 
Summing over proton and helium contributions, we obtain the total BDM flux
\begin{align}
\label{eq:fluxBDM}
    \frac{d\Phi_\mathrm{\chi}}{d \Tx} = D_\mathrm{eff} \frac{\rho_\chi}{\mx} \sum_{i} G_i^2(q^2) \int_{\Ti^{\mathrm{min}}}^\infty \frac{d\Phi_\mathrm{\mathrm{LIS}}}{d \Ti} \frac{d \sigma_{\chi i}}{d \Tx} d\Ti,
\end{align}
shown in figure \ref{fig:fluxBDMNucleons} for the case of energy independent cross section (blue), and the four heavy mediator scenarios discussed above for fermion DM.

\section{Boosted Dark matter scattering in the detector}
\subsection{Event rate}
The BDM particles scattered towards the detectors carry enough energy to induce a signal in the target material. The natural reference frame for this process is the laboratory frame, where the target nucleus is at rest and the boosted DM particle has a large velocity. Hence, the cross sections can be translated from CR-DM scattering processes to BDM-target nucleus scattering using the transformations $\mx \rightarrow m_t$, $m_i\rightarrow \mx$, $\Tx \rightarrow E_\mathrm{R}$, $\Ti \rightarrow \Tx$. Here, $E_\mathrm{R}$ and $m_t$ are the recoil energy and mass of the target nucleus in the detector. This consequently transforms $\Tx^{\mathrm{max}} \rightarrow E_\mathrm{R}^\mathrm{max}$ and likewise for the center of mass energy squared in equation~\eqref{eq:MandelstamS} \cite{Ghosh:2024dqw}. 

The differential BDM scattering rate with respect to the recoil energy is thus obtained as
\begin{align}
\label{eq:diffRate}
    \frac{dR}{d E_\mathrm{R}} = \sum_{t} \epsilon_t N_t G_t^2(q^2) \int_{\Tx^{\mathrm{min}}}^\infty \frac{d\Phi_\mathrm{\chi}}{d \Tx} \frac{d \sigma_{\chi t}}{d \Er} d\Tx.
\end{align}
Here, $\epsilon_t$ denotes the detection efficiency for recoils off the target nucleus $t$, $d\Phi_\mathrm{\chi}/d \Tx$ the BDM flux given by equation (\ref{eq:fluxBDM}) and $N_t$ the number of target particles in the detector. The remaining terms follow applying the transformations discussed above to equation~\eqref{eq:fluxBDM}. For the nuclear form factor $G_t^2$ we use the Helm form factor, including the term $A_t^2 \mu_{\chi t}/\mu_{\chi n}$ as in equation \eqref{eq:formfactor}.

\subsection{Attenuation effects}
The COSINUS experiment is located within the Laboratori Nazionali del Gran Sasso, a deep underground laboratory. In the standard DM nuclear recoil searches, probing DM in the MeV -- TeV regime, the direct detection experiments are sensitive to such small DM-nucleon cross sections that attenuation effects are negligible. However, in the case of BDM and light DM masses, the cross sections that are within experimental reach can be large enough for relevant attenuation effects to take place. These effects reduce the DM particle's kinetic energy, hence reducing the event rate above threshold in the detectors. We use the depth $z = \SI{1.4}{\km}$ and composition of Gran Sasso reported in \cite{alveyNoRoomHide2023}. The attenuation effects can be determined solving the differential equation \cite{bellCosmicrayDarkMatter2024, alveyNoRoomHide2023, maityCosmicrayBoostedDark2024}
\begin{align}
    \frac{d\Tx^z}{dz} = -\sum_N n_N \int_0^{\Er^{\mathrm{max}}} \Er \frac{d \sigma_{\chi N}}{d\Er} d\Er,
\end{align}
where $n_N$ labels the average density of an isotope $N$ in the overburden. Notice, depending on the DM candidate, the differential cross section can have dependencies on $\Tx$ and $\Er$. The differential equation can be solved numerically obtaining $\Tx^z$, the attenuated DM kinetic energy at depth $z$. We can then invert the solution to obtain the initial kinetic energy as a function of the attenuated kinetic energy, $\Tx(\Tx^z)$. Finally, the attenuated DM differential flux is obtained via the chain rule as
\begin{align}
    \frac{d\Phi_\mathrm{\chi}}{d \Tx^z} = \frac{d\Tx}{d\Tx^z} \frac{d\Phi_\mathrm{\chi}}{d \Tx}.
\end{align}
Using this numerical procedure we found that the effects are negligible for our specific choice of exposure. Thus, the results reported later are not including this effect to simplify the calculations and cut down on computational costs. 

\section{Statistical Evaluation}
\label{sec:StatEval}

In this work, we report expected bounds in the sub-GeV DM regime for COSINUS given a likelihood analysis, which is reported in terms of the DM-nucleon reference cross section. The definition of the cross section can be found in Appendix \ref{app:cs-formalism}. Our projection is based on exposure of $\SI{100}{kg\,d}$.

We find the expected limit using a Monte Carlo procedure: Starting from a  simulation based background model for COSINUS \cite{angloherCOSINUSModelindependentSensitivity2025, angloherCOSINUSModelindependentChallenge2025, COSINUS:2021bdj} we draw a sample of random variates from a Poisson distribution for each energy bin. In this work we assume 50 nuclear recoil energy bins for COSINUS with a constant bin width of $\SI{1}{keV}$. The nuclear recoil energy region of interest is taken to extend from $\SI{1}{keV}$ to $\SI{51}{keV}$.% 

The free parameters in the model are the DM mass $\mx$ and the DM-nucleon reference cross section $\bar\sigma_{\chi i}$. We perform the analysis on a grid of points in the DM mass, restoring the problem to a one-dimensional search for a cross section limit for each value of the DM mass.
The likelihood function is given as
\begin{align}
    \mathcal{L}(\bar\sigma_{\chi i}) = \prod_{j} e^{-(b_j + S_{j}(\bar\sigma_{\chi i}))} \frac{(b_j + S_{j}(\bar\sigma_{\chi i}))^{n_j}}{n_j!},
\end{align}
where $S_{j}(\bar\sigma_{\chi i})$ denotes the expected signal in the $j$-th bin, obtained by integrating the differential rate given in equation \eqref{eq:diffRate} over the energy bin, folded with a Gaussian smearing kernel of width given by the energy resolution of the detector. Here we assume a 200 eV resolution, but our results are not sensitive to the energy resolution in the recoil energy region above a few keV. The expected number of  background events is given by $b_j$ and the randomly drawn number of events as $n_j$ in the $j$-th bin. Our null hypothesis is a background-only model, corresponding to $\bar\sigma_{\chi i}=0$. The likelihood ratio $\lambda$ and test statistic $q$ are defined as~\cite{cowanAsymptoticFormulaeLikelihoodbased2011}
\begin{align}
    \lambda(\bar\sigma_{\chi i}) = \frac{\mathcal{L}(\bar\sigma_{\chi i})}{\mathcal{L}(0)}, \quad q(\bar\sigma_{\chi i}) = - 2 \log(\lambda(\bar\sigma_{\chi i})).
\end{align}
We determine the exclusion limit by finding the value of the reference cross section $\bar\sigma_{\chi i}$ for which the test statistic equals $q_0 = 1.645^2$, corresponding to a 90\% confidence level exclusion of the signal hypothesis. We repeat this procedure for $N=10^5$ times, each with a new random sample of the event counts $n_j$, resulting in a sample of exclusion limits. From this sample we find the median expected exclusion limit and the 90\% confidence level range for the exclusion limit.

\section{Results}

We will now utilize the statistical procedure described above to infer expected limits for fermion, scalar or vector dark matter, considering the energy independent interaction as well as interactions mediated by a scalar, pseudoscalar, vector or axial vector boson. We will first outline the formalism for obtaining the energy dependent differential cross sections, following similar conventions as e.g. in \cite{bardhanBoundsBoostedDark2023}. A more detailed discussion is presented in appendix \ref{app:cs-formalism}. 

The constrains set by the energy independent interaction are conservative estimates for the BDM sensitivity. Introducing energy dependent interactions can improve the sensitivity by orders of magnitude, but introduces model-dependence. Generally, the free parameters of the interaction model would also include the mass of the mediating particle. However, here we work on the limit of heavy mediator mass, $m_\phi^2\gg q^2$, to reduce the number of free parameters and to allow simpler comparison between the energy independent and the various energy dependent scenarios. We note that a light mediator particle would additionally be subject to various constraints from cosmology and collider experiments \cite{Brust:2017nmv,bellCosmicrayDarkMatter2024}, which we avoid in the heavy mediator limit.  We note that even in the heavy mediator case, constraints e.g. from big bang nucleosynthesis generally apply \cite{Krnjaic:2019dzc}. however, translating these bounds into DM-nucleon elastic scattering cross section depends on details of the DM model and on the cosmic history. Hence, we do not impose these constraints on the parameter space studied below.

All of the DM-nucleon scattering processes considered in this work are described by a $t$-channel diagram. We define a scattering function $\mathcal{K}$ via the relation
\begin{equation}
    |\mathcal{M}|^2=\frac{\bar\sigma_{\chi i} \pi}{\mu_{\chi i}^2}\mathcal{K}(s,t),
\label{eq:K-function}
\end{equation}
where $\mathcal{M}$ is the scattering amplitude and the reference DM-nucleon cross section $\sigma_{\chi i}$ is given by
\begin{equation}
    \bar\sigma_{\chi i} = \frac{\lambda_\chi^2 \lambda_i^2 \mu_{\chi i}^2}{\pi \left(q^2_\mathrm{ref} - m_\phi^2\right)^2},
    \label{eq:referencexsec}
\end{equation}
where $\lambda_\chi$, $\lambda_i$ are the DM-mediator and mediator-nucleon couplings, $m_\phi$ is the mediator mass and $q^2_\mathrm{ref}$ the reference momentum transfer squared. With this convention, the reference cross section corresponds to the total cross section obtained from the scattering amplitude in the limit of large mediator mass and zero momentum transfer. As shown in appendix \ref{app:cs-formalism}, we can then obtain the differential cross section $d\sigma_{\chi i}/dT_\chi$ in terms of the scattering function $\mathcal{K}$, and substitute this to equations (\ref{eq:fluxBDM}) and (\ref{eq:diffRate}) to obtain the DM event rate for each model.

The projected limits reported in this work are obtained using an assumed exposure of \SI{100}{kg\,d} in ${\rm NaI}$ and the estimated background rate used in \cite{angloherCOSINUSModelindependentSensitivity2025}. Moreover, we assume a signal efficiency of $\epsilon_\mathrm{Na} =0.38$ for Sodium recoils and $\epsilon_\mathrm{I} = 0.76$ for Iodine recoils \cite{angloherCOSINUSModelindependentChallenge2025}.

\subsection{Fermion Dark matter constrains}

\begin{figure}[tb]
    \centering
    \includegraphics[width=.75\linewidth]{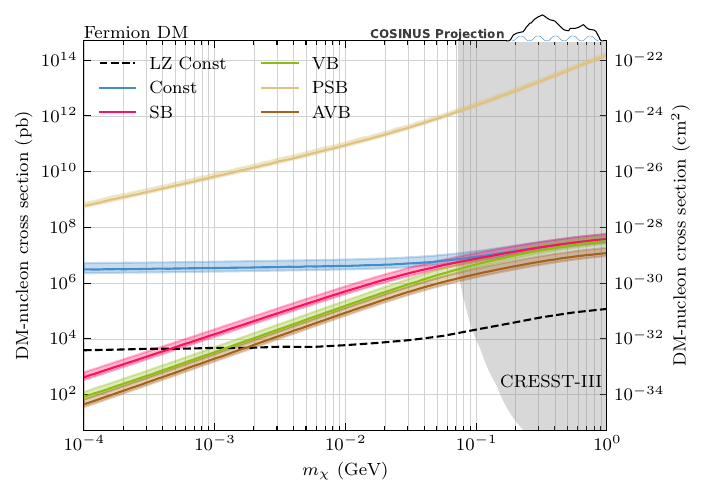}
    \caption{Expected light fermion DM bounds for COSINUS with 100 kg d exposure for energy independent (blue) and heavy mediator interactions. The latter is for scalar (red), vector (green), pseudoscalar (beige) and axial vector (brown) mediators, respectively. The shaded areas show the 90\% confidence level uncertainty in the expected limit, as explained in section \ref{sec:StatEval}. The results are portrayed in comparison to the strongest BDM results, obtained by LZ \cite{LZ:2025iaw}. In addition, we include the limits obtained by CRESST-III \cite{CRESST:2024cpr}.}
    \label{fig:FMComparison}
\end{figure}

Figure \ref{fig:FMComparison} presents the results obtained with the cross section formulas (\ref{eq:FMSBCrossSection}) - (\ref{eq:FMAVBCrossSection}). The lines shown are the 90\% confidence level exclusion limits, with the 90\% uncertainty bands. We notice that apart from the pseudoscalar mediator case, every other scenario is a large improvement in sensitivity compared to the energy-independent model. Therefore, the energy-independent model can be taken as a conservative limit, with the caveat of the pseudoscalar mediator, which is known to be poorly constrained in direct detection due to the momentum transfer suppressed cross section. For the purpose of constraining a specific model, it is of course advisable to use the correct energy dependent interaction cross section, as it can lead to a significant increase in sensitivity.  

\subsection{Scalar Dark matter constrains}

As described in appendix \ref{app:sb-calc}, the differential cross sections for scalar dark matter interacting via (pseudo)scalar or (axial)vector mediator are given by
\begin{align}
    &\left(\frac{d \sigma_{\chi i}}{d \Tx}\right)_{\mathrm{SB, SB}} = \frac{\bar\sigma_{\chi i} \mx^2}{\mu^2 s T^\mathrm{max}_\chi} \left(\mi^2 + \frac 12 \mx \Tx \right), \\ %END
    &\left(\frac{d \sigma_{\chi i}}{d \Tx}\right)_{\mathrm{SB, VB}} = \frac{\bar\sigma_{\chi i} \mx^2}{2\mu^2 s T^\mathrm{max}_\chi} \Bigl[2(\mi + \Ti)^2 - \Tx(2\mi + 2\Ti + \mx)\Bigr], \\
    &\left(\frac{d \sigma_{\chi i}}{d \Tx}\right)_{\mathrm{SB, PSB}} = \frac{\bar\sigma_{\chi i} \mx^3 \Tx}{32\mu^2 s T^\mathrm{max}_\chi}, \\ %END
    &\left(\frac{d \sigma_{\chi i}}{d \Tx}\right)_{\mathrm{SB, AVB}} = \frac{\bar\sigma_{\chi i} \mx}{2\mu^2 s T^\mathrm{max}_\chi} \Bigl[2 \mx \Ti \left(2 \mi + \Ti\right) - \Tx \left(2 \mx \Ti\right) + \left(\mi+ \mx\right)^2\Bigr].
\end{align}
The Feynman rules we've adopted for obtaining these formulas are detailed in appendix \ref{app:sb-calc}. 

\begin{figure}[tb]
    \centering
    \includegraphics[width=.75\linewidth]{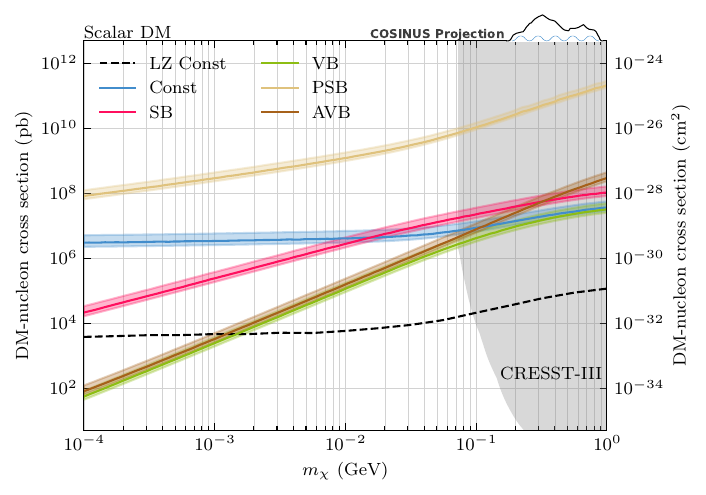}
    \caption{Expected light scalar DM bounds for COSINUS with 100 kg d exposure for energy independent (blue) and heavy mediator interactions. The latter is for scalar (red), vector (green), pseudoscalar (beige) and axial vector (brown) mediators, respectively. The shaded areas show the 90\%  uncertainty in the expected limit. The results are portrayed in comparison to the strongest BDM results, obtained by LZ \cite{LZ:2025iaw}. In addition, we include the limits obtained by CRESST-III \cite{CRESST:2024cpr}.}
    \label{fig:SBComparison}
\end{figure}

Similarly to the fermion DM case, figure \ref{fig:SBComparison} portrays the results for scalar DM. 
We find that the pseudoscalar case again leads to momentum transfer suppressed event rates and therefore weaker limits. Compared to fermion DM, however, here the suppression is only linear in $T_\chi$, instead of scaling as $T_\chi^2$. 

Contrary to earlier, the vector mediator case provides slightly more stringent bounds than the axial vector case.

\subsection{Vector Dark matter constrains}
Similarly to the previous sections, we can obtain the differential cross sections for the case of vector DM. A scenario in which vector DM couples to the Standard Model (SM) through SM photons has been explored in \cite{hisanoDirectDetectionVector2020}. In that framework, the DM particles are charged under a dark $SU(2)$ gauge symmetry, and two additional dark fermions are introduced to induce a loop-level coupling between the DM particle and the SM photon. In a model introduced by \cite{belyaevFermionicPortalVector2023} the fermion loop couples instead to the $Z$ boson, thereby providing a massive mediator coupling to the SM.
In this work we do not specify the quantum numbers of the mediator, but simply assume the interaction to be mediated by an unspecified heavy vector boson. We model this interaction with a three-vector vertex, which in a UV-complete theory could arise e.g. from integrating out a heavy fermion or scalar coupling to both vector bosons. For the case of a scalar mediator, we adopt a vertex rule analogous to the Higgs-$Z$ coupling. Similarly, we adopt the calculations for the pseudoscalar and axial vector case. The explicit Feynman rules used in our calculations can be found in appendix \ref{app:vb-calc}.

\begin{figure}[tb]
    \centering
    \includegraphics[width=.75\linewidth]{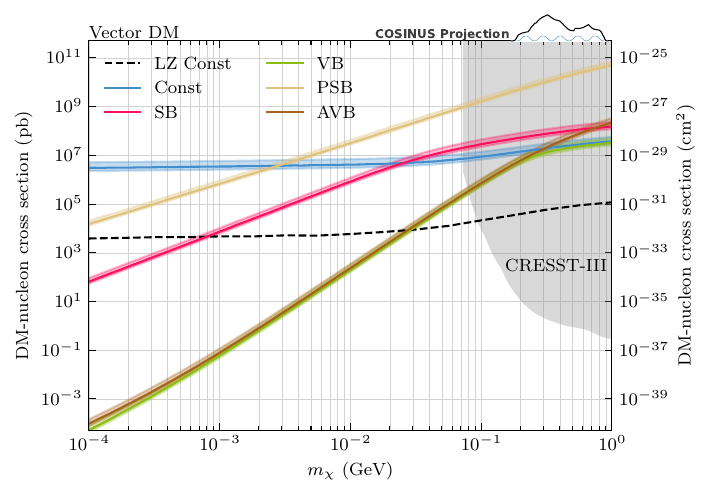}
    \caption{Expected light vector DM bounds for COSINUS with 100 kg d exposure for energy independent (blue) and heavy mediator interactions. The latter is for scalar (red), vector (green), pseudoscalar (beige) and axial vector (brown) mediators, respectively. The shaded areas show the 90\%  uncertainty in the expected limit. The results are portrayed in comparison to the strongest BDM results, obtained by LZ \cite{LZ:2025iaw}. In addition, we include the limits obtained by CRESST-III \cite{CRESST:2024cpr}.}
    \label{fig:VBComparison}
\end{figure}

Under these assumptions, we obtain the differential cross sections for vector DM as given in equations \eqref{eq:VDMSB}-\eqref{eq:VDMAVB}. The resulting projections on the COSINUS sensitivity to vector DM are presented in figure \ref{fig:VBComparison}.

\begin{align}
    \left(\frac{d \sigma_{\chi i}}{d \Tx}\right)_{\mathrm{VB, SB}}=&\frac{\bar\sigma_{\chi i}}{24\mu^2 s T^\mathrm{max}_\chi} \left(2\mx\Tx+3\mx^2+\Tx^2\right)\left(2\mi^2+\mx\Tx\right), \label{eq:VDMSB} \\%END
    \left(\frac{d \sigma_{\chi i}}{d \Tx}\right)_{\mathrm{VB, VB}}= &\frac{\bar\sigma_{\chi i}}{6\mu^2 s T^\mathrm{max}_\chi} \Bigr[T_\chi^3 \left(-2 m_i-2 T_i+3 m_\chi\right) +2 T_\chi^2 \left(2 m_i \left(T_i-m_\chi\right)-2 T_i m_\chi-m_i^2+T_i^2+3
    m_\chi^2\right)\nonumber\\
    &-m_\chi T_\chi \left(m_i \left(6 m_\chi-8 T_i\right)+6 T_i m_\chi+4 m_i^2-4 T_i^2+3 m_\chi^2\right)+6 m_\chi^2 \left(m_i+T_i\right){}^2\Bigr], \\%END
     \left(\frac{d \sigma_{\chi i}}{d \Tx}\right)_{\mathrm{VB, PSB}}=&\frac{\bar\sigma_{\chi i} \mx  \Tx}{24\mu^2 s T^\mathrm{max}_\chi} \left(2\mx\Tx+3\mx^2+\Tx^2\right), \\%END
%    \end{align}
%    \begin{align}
    \left(\frac{d \sigma_{\chi i}}{d \Tx}\right)_{\mathrm{VB, AVB}}= &\frac{\bar\sigma_{\chi i}}{6\mu^2 s T^\mathrm{max}_\chi \mx} \Bigl[\mi^2 \Tx \left(6\mx \Tx + 13 \mx^2 - \Tx^2\right) + 2 \mi \mx \left(2\Ti - \Tx\right)\left(2\mx\Tx + 3\mx^2 + \Tx^2\right) \nonumber\\
    &+ \mx\left(6\mx^2\left(\Tx^2+\Ti^2-\Ti\Tx\right) + \mx \Tx \left(4\Ti^2 + 3\Tx^2 - 4\Ti\Tx\right) + 2 \Ti \Tx^2 \left(\Ti - \Tx\right) - \mx^3 \Tx\right)\Bigr]\label{eq:VDMAVB}
\end{align}

\subsection{Boosted Dark matter from DSNB}
The results above are based on a single effective interaction cross section between the DM particle and the nucleons, $\bar\sigma_{\chi i}$. However, if the DM particle additionally has interactions with neutrinos, there will be an additional source of boosted DM flux from the cosmic high energy neutrino background. 
Although interactions with neutrinos would in many concrete model frameworks also give rise to DM-electron interactions, in this work we do not consider DM-electron scattering and only include DM-nucleon interactions in the detector. We note that any electron scattering events in the detector could only increase the total signal rate and hence our limits that ignore the electron scattering are conservative. As the neutrino source, we assume a diffuse supernova neutrino background (DSNB). While this has not been detected yet \cite{Abe:2025qug}, we make use of neutrino flux models based on theory and simulations. In order to obtain a neutrino boosted DM flux, we determine the flavor dependent neutrino flux with respect to the redshift $z$, 
\begin{align}
    \frac{d \Phi_\nu^\alpha}{d T_\nu} = \int_0^{z_{\mathrm{max}}} \frac{R_\mathrm{CCSN}}{H(z)} F^\alpha_\nu([1+z] T_\nu)\,dz,
\end{align}
where $z_\mathrm{max} \approx 6$ \cite{Ghosh:2021vkt, Ghosh:2024dqw}. The energy of the neutrino is denoted by $T_\nu$ and $\alpha$ depicts the flavor of the neutrino. Moreover, the equation contains the rate of core collapse supernova (CCSN) $R_\mathrm{CCSN}$, the Hubble parameter $H(z)$ and the Fermi-Dirac distribution $F^\alpha_\nu(T_\nu)$. The latter two are given as
\begin{align}
    &H(z) = H_0 \sqrt{\Omega_\Lambda + \Omega_m (1 + z)^3}, \\
    &F^\alpha_\nu(T_\nu) = \frac{120 T_\nu^{tot} T_\nu}{42 \pi^4 t_\alpha^4} \frac{1}{e^{T_\nu/t_\alpha} + 1},
\end{align}
where the Hubble constant $H_0 = \SI{67.2}{km.s^{-1}.Mpc^{-1}}$, the dark energy parameter $\Omega_\Lambda = 0.68$, the matter density $\Omega_\mathrm{m} = 0.3$ \cite{collaboration_planck_2020}. The parameters of the neutrino Fermi-Dirac distribution are the total neutrino energy emitted from supernovas $T_\nu^{\mathrm{tot}} \approx \SI{3e55}{GeV}$ and the temperature $t_\alpha$ of each neutrino flavor $\alpha$. The temperatures are given as $t_{\overline{\nu}_e} = \SI{7}{MeV}$, $t_{\nu_e} = \SI{6.6}{MeV}$, and $t_{\nu_{\tau/\mu}} = \SI{10}{MeV}$. The CCSN rate is expressed by
\begin{align}
    R_\mathrm{CCSN}(z) &= \dot{\rho}_0 \left[ (1+z)^{\alpha\eta} + \left( \frac{1+z}{B} \right)^{\beta\eta} + \left( \frac{1+z}{C} \right)^{\gamma\eta} \right]^{1/\eta} \SI{0.0109}{M_\odot^{-1}},\\
    B &= (1+z_1)^{1-\alpha/\beta},\\
    C &= (1+z_1)^{(\beta-\alpha)/\gamma}(1+z_2)^{1-\beta/\gamma},
\end{align}
where $\eta = -10$, $\alpha =3.4$, $\beta=-0.3$ and $\gamma=-3.5$ are fit parameters of an empirical model. The normalization is given as $\dot{\rho}_0 = 0.0178$. The mass $\SI{0.0109}{M_\odot^{-1}}$ is determined by the ratio of the integrals over the initial mass functions \cite{DeGouvea:2020ang}. Further, the redshift breaks are given by $z_1 = 1$ and $z_2 = 4$.

\begin{figure}
    \centering
    \includegraphics[width=.7\linewidth]{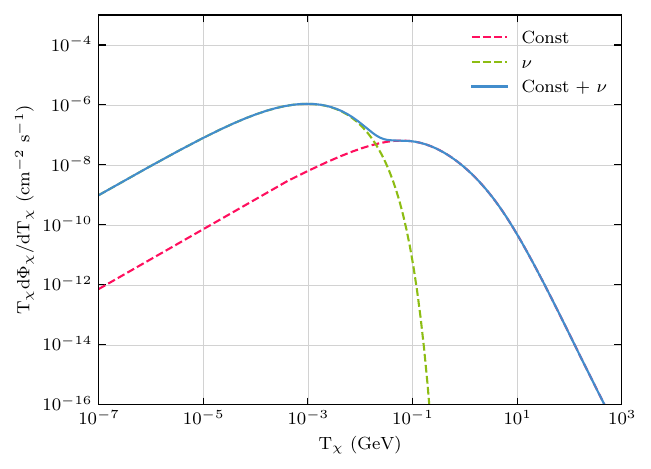}
    \caption{Same as figure \ref{fig:fluxBDMNucleons} but in addition with DSNB contributions, labelled by $\nu$. The cross section is chosen to be $\bar\sigma_{\chi i} = \bar\sigma_{\chi \nu} = 10^{-30}\,\mathrm{cm}^2$ and the DM mass $m_\chi = \SI{0.1}{GeV}$.}
    \label{fig:fluxBDMNeutrino}
\end{figure}

\begin{figure}
    \centering
    \includegraphics[width=.75\linewidth]{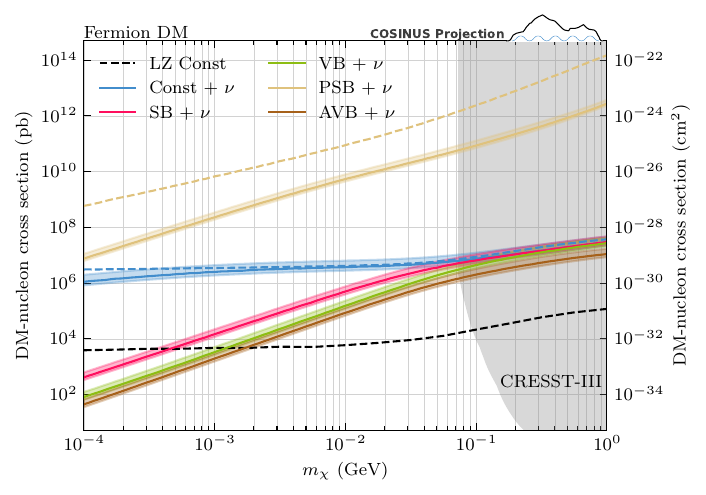}
    \caption{Expected light fermion DM bounds for COSINUS with 100 kg d exposure for energy independent (blue) and heavy mediator interactions with nuclei, including an energy independent interaction with neutrinos, where the reference cross section with neutrinos is taken equal to the cross section with nuclei. The mediator cases depict the same as previously. The solid curves portray the processes including neutrinos, while the dashed curves correspond to the previously shown neutrinoless scattering processes. Including neutrinos led to a significant improvement solely in the energy independent and in the pseudoscalar case. In comparison, we again portray the limits obtained by LZ and CRESST-III \cite{CRESST:2024cpr, LZ:2025iaw}.}
    \label{fig:FMComparisonNeutrino}
\end{figure}

\begin{figure}[tb]
    \centering
    \includegraphics[width=.75\linewidth]{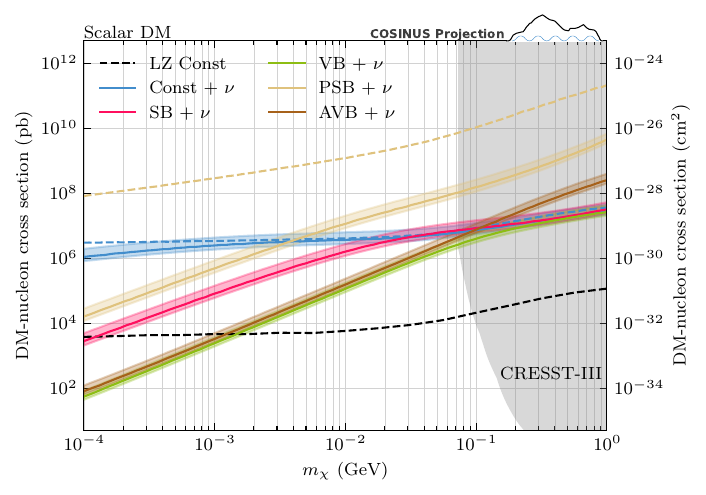}
    \caption{Same as figure \ref{fig:FMComparisonNeutrino} for scalar DM.}
    \label{fig:SBComparisonNeutrino}
\end{figure}

\begin{figure}[tb]
    \centering
    \includegraphics[width=.75\linewidth]{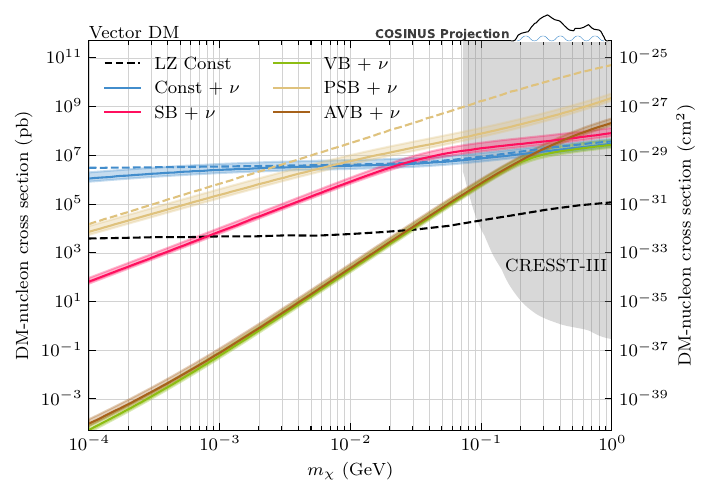}
    \caption{Same as figure \ref{fig:FMComparisonNeutrino} for vector DM.}
    \label{fig:VBComparisonNeutrino}
\end{figure}

Depending on the UV-complete model for DM-neutrino interactions, the coupling could be flavor dependent.
However, for simplicity, we demonstrate the boost due to neutrinos assuming the same coupling strength for each flavor. In addition, we assume the same coupling strength for DM with neutrinos and nucleons, $\bar{\sigma}_{\chi i} = \bar{\sigma}_{\chi \nu}$. Last, we assume an energy independent interaction between DM and neutrinos,
\begin{align}
    \frac{d\sigma_{\chi \nu}}{d T_\chi} = 
    \frac{\bar{\sigma}_{\chi \nu}}{T_\chi^{\max}\left(T_\nu\right)}\,
    \Theta\!\left(T_\chi^{\max}\left(T_\nu\right) - T_\chi\right),
\end{align}
In analogy to the energy independent DM-nucleon interaction introduced earlier. The maximum kinetic energy follows the same principle as above under the assumption that the neutrino is massless \cite{Ghosh:2021vkt, Ghosh:2024dqw}
\begin{align}
    T_\chi^{\mathrm{max}} = \frac{T_\nu^2}{T_\nu + \frac12 m_\chi}.
\end{align}
Similarly, we obtain the minimum required energy to obtain a BDM particle of energy $T_\chi$ using equation \eqref{eq:TiMin} under the assumption of massless neutrinos. The BDM flux due to the DSNB is then given as
\begin{align}
    \frac{d\Phi_\mathrm{\chi}}{d \Tx} = D_\mathrm{eff} \frac{\rho_\chi}{\mx} \sum_\alpha\int_{T_\nu^{\mathrm{min}}}^\infty \frac{d\Phi_\nu^\alpha}{d T_\nu} \frac{\bar{\sigma}_{\chi \nu}}{T_\nu^{\max}} dT_\nu.
\end{align}
Finally, the total BDM flux obtained through scattering off neutrinos and CR particles is the sum over all single contributions
\begin{align}
    \frac{d \Phi^{tot}_\chi}{d T_\chi} = \sum_{i=p, ^4\mathrm{He}, \nu} \left(\frac{d\Phi_\chi}{d\Tx}\right)_i,
\end{align}
shown in figure \ref{fig:fluxBDMNeutrino}. We observe that the DSNB scattering adds a second maximum to the BDM flux at a lower energy compared to the CR scattering. The expected 90\% CL exclusion limits for $\bar{\sigma}_{\chi i} = \bar{\sigma}_{\chi \nu}$ are shown in figures \ref{fig:FMComparisonNeutrino}, \ref{fig:SBComparisonNeutrino} and \ref{fig:VBComparisonNeutrino} for fermion, scalar and vector DM.

\section{Conclusion}
In this work, we explored the potential of BDM to extend the sensitivity of COSINUS to light dark matter. We focused on CR particles scattering off non-relativistic DM particles in the halo, providing higher kinetic energies to DM and thus enabling detection in COSINUS. 

We described the DM-nucleon interactions with an energy independent cross section, and a set of energy dependent scenarios corresponding to integrating out a heavy mediator that is either a scalar, pseudoscalar, vector or an axial vector. Furthermore, we carried out the analysis for fermion, scalar and vector DM. We find that generally the pseudoscalar mediator scenario is the least constrained due to the momentum transfer suppressed cross sections, followed by the energy independent case. All other heavy mediator scenarios lead to much improved sensitivity compared to these two cases. We have also considered possible DM-neutrino interactions, and shown how the boosted DM flux due to the DSNB can improve the sensitivity to light DM.

\section*{Acknowledgments}
The COSINUS collaboration thanks the directors and the staff of the Laboratori Nazionali del Gran Sasso and the technical staff of our laboratories. Furthermore, the collaboration expresses its gratitude to Alejandro Ibarra and Yu Chen for their earlier contributions in supplying initial input on this topic. This work has been supported by the Max-Planck-Institut für Physik, the Marietta-Blau-Institut für Teilchenphysik der Österreichischen Akademie der Wissenschaften, the Technische Universität Wien, the Istituto Nazionale di Fisica Nucleare and the Helsinki Institute of Physics. Financial support from the Research Council of Finland (grant$\#$ 371542), the Finnish Graduate School in Particle and Nuclear Physics (Doctoral Education Pilot), Vilho, Yrjö and Kalle Väisälä fund, Austrian Science Fund FWF (grant DOI 10.55776/PAT1239524) and Klaus Tschira Foundation is gratefully acknowledged. The authors wish to thank the Finnish Computing Competence Infrastructure (FCCI) for supporting this project with computational and data storage resources. HILE is funded by ERC grant (ILLUMINATOR, 101114623).

%%%%%%%%%%%%%%%%%%%%%%%%%%%%%%%%%%%%%%%%%%%%%%%%
\bibliography{refs.bib}
%%%%%%%%%%%%%%%%%%%%%%%%%%%%%%%%%%%%%%%%%%%%%%%%%%
%%%%%%%%%%%%%%%%%%%%%%%%%%%%%%%%%%%%%%%%%%%%%%%%
%%%%%%%%%%%%%%%%%%%%%%%%%%%%%%%%%%%%%%%%%%%%%%%%

\newcommand{\mprod}[2]{p_#1 \cdot p_#2}
\newcommand{\up}[1]{\mathrm{u(p_{#1})}}
\newcommand{\upb}[1]{\mathrm{\bar{u}(p_#1)}}
\newcommand{\denom}[1]{q^2 - m_{#1}^2 + i\epsilon}
\newcommand{\denomSqr}[1]{(q^2 - m_{#1}^2)^2}
\newcommand{\dcs}{\frac{\mathrm{d}\sigma}{\mathrm{d T_\chi}}}

\newpage
\appendix
\section{Cross section formalism} \label{app:cs-formalism}
As discussed in the main article, the cross section calculations are naturally performed in one of the two frames: the galactic rest frame for the cosmic ray scattering off DM, where the DM particle is initially taken to be at rest, or the laboratory frame for the detector, where the target nucleus is initially taken to be at rest.

The Lorentz invariant differential cross section is given by
\begin{equation}
\label{eq:dsigmadt}
    \frac{d\sigma}{dt} = \frac{\left|\mathcal{M}\right|^2}{16 \pi \lambda(s, m_\chi^2, m_i^2)},
\end{equation}
where the Mandelstam variable $s$ is the center of mass energy squared.

The function in the denominator is the Källén function
\begin{equation}
    \lambda(x, y, z) = x^2 + y^2 + z^2 -2xy -2xz -2yz.
\end{equation}
Further, we define a reference cross section to obtain a comparable quantity between all differential cross sections, given by
\begin{align}
    \bar\sigma_{\chi i} = \frac{\lambda_\chi^2 \lambda_i^2 \mu_{\chi i}^2}{\pi \left(q_\mathrm{ref}^2 - m_\alpha^2\right)^2}.
\end{align}
Here $\lambda_\chi$ and $\lambda_i$ are dimensionless coupling constants between the mediator and the DM and SM particle, respectively,  $q_\mathrm{ref}$ is the reference momentum transfer, $m_\alpha$ the mediator mass and $\mu_{\chi i}$ the reduced mass of the DM and SM particles. 
The reference cross section is related to the scattering amplitude squared via
\begin{align}
\label{eq:Msquared}
    \left|\mathcal{M}\right|^2 = \frac{\bar\sigma_{\chi i} \pi}{\mu_{\chi i}^2} \mathcal{K}(s, t),
\end{align}
where the scattering function $\mathcal{K}(s,t)$ describes the energy dependence of the scattering amplitude in terms of the Mandelstam variables $s$ and $t$. Substituting equation \eqref{eq:Msquared} into equation \eqref{eq:dsigmadt} yields an expression for the differential cross section in terms of the reference cross section
\begin{align}
    \frac{d\sigma}{dt} = \frac{\bar\sigma_{\chi i} \mathcal{K}(s, t)}{16 \mu_{\chi i}^2\lambda(s, m_\chi^2, m_i^2)}.
\end{align}
Last, we will express this Lorentz-invariant quantity in the galactic or laboratory frame in terms of the kinetic energies. The frame dependent energy is denoted as $T_k$ for a final state particle with mass $m_k$. The differential cross section in the galactic or laboratory frame is then given by
\begin{align}
    \frac{dt}{dT_k} =& - 2 m_k, \\
    \frac{d\sigma}{dT_k} =& \left| \frac{dt}{dT_k} \right| \frac{d\sigma}{dt}.
\end{align}
In the fixed frame the Källén function simplifies to
\begin{equation}
    \lambda(s, m_k^2, m_p^2) = 2 m_k T_k^{\mathrm{max}} s.
\end{equation}
Finally, putting all the steps together, we get the galactic ($k=\chi$) or lab ($k = i$) dependent cross section,
\begin{align}
    \frac{d\sigma}{d T_k} = \frac{\bar\sigma_{\chi i} \mathcal{K}(s, t)}{16 \mu_{\chi i}^2 s T_k^{\mathrm{max}} }.
\end{align}

\section{Fermion DM}
\label{Appendix:fermion}
For fermion DM with a scalar mediator we use the vertex rule

\begin{figure}[H]
  \centering
  $\vcenter{\hbox{\includegraphics[scale=0.8]{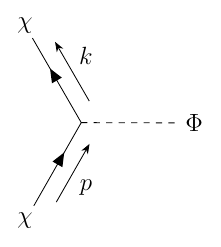}}}$
  $\displaystyle = - i \lambda_\chi$,
\end{figure}

\noindent and similarly for the mediator-nucleon vertex, given by the coupling $\lambda_i$. Evaluating the squared scattering amplitude we find
\begin{align}
    \mathcal{K}(s, t) = 4 \left(2 m_k^2 - \frac t2\right) \left(2 m_p^2 - \frac t2\right).
\end{align}

\noindent For the coupling with the vector mediator we take

\begin{figure}[H]
  \centering
  $\vcenter{\hbox{\includegraphics[scale=0.8]{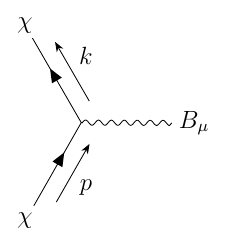}}}$
  $\displaystyle = - i \lambda_\chi \gamma^\mu $.
\end{figure}

\noindent Evaluating the squared scattering amplitude yields
\begin{equation}
    \mathcal{K}(s, t) = 2\bigl[2 s^2+2 s t+t^2  +
    \left(m_k^2+m_{p}^2\right) \left(m_k^2+m_{p}^2-2 s+2 t\right)\bigr].
\end{equation}

\noindent Exchanging the mediators through their parity-reversed counterparts, the pseudoscalar $\eta$ and axial vector $A_\mu$, adds a factor of $\gamma^5$ to the vertices including fermions.

Hence, we obtain the following vertices and corresponding scattering functions.
\begin{figure}[H]
  \centering
  $\vcenter{\hbox{\includegraphics[scale=0.8]{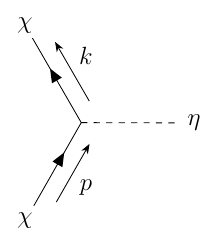}}}$
  $\displaystyle = - i \lambda_\chi \gamma^5$,
\end{figure}

yielding
\begin{align}
    \mathcal{K}(s, t) = t^2.
\end{align}

And for the axial vector mediator
\begin{figure}[H]
  \centering
  $\vcenter{\hbox{\includegraphics[scale=0.8]{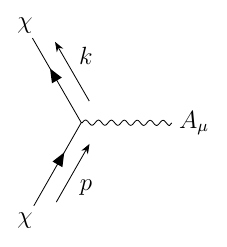}}}$
  $\displaystyle = - i \lambda_\chi \gamma^\mu \gamma^5$,
\end{figure}
yielding
\begin{align}
    \mathcal{K}(s, t) = \frac{1}{8}\Bigl[2\bigl(m_k^4 + m_p^4 - 2 m_p^2 (s + t) - 2 m_k^2(s+t- 5m_p)\bigr) + 2 s^2 + 2st + t^2\Bigr].
\end{align}

\section{Scalar DM} \label{app:sb-calc}
For scalar DM and scalar mediator we take the three scalar vertex rule as
\begin{figure}[H]
  \centering
  $\vcenter{\hbox{\includegraphics[scale=0.8]{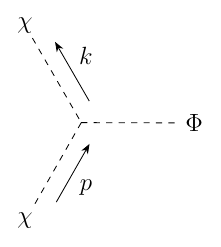}}}$
  $\displaystyle =  -\frac i2 \lambda_\chi m_\chi$.
\end{figure}

\noindent The scalar three-point function has dimension of mass, and we parametrize this with the DM mass $m_\chi$. In a UV-complete theory the DM mass may or may not be associated with the mass scale appearing in the vertex rule, but one can always recover our parametrization by redefinition of the coupling constant $\lambda_\chi$. The resulting scattering function is given by
\begin{align}
    \mathcal{K}(s, t) = m_\chi^2 \left(m_k^2 - \frac t4\right).
\end{align}

\noindent In the same fashion, we can determine the pseudoscalar case under the appropriate vertex transformations of the fermionic scattering partners,
\begin{align}
    \mathcal{K}(s, t) = -\frac{m_\chi^2 t}{4}.
\end{align}
\noindent For the coupling between the scalar DM and vector mediator we use the Feynman rule

\begin{figure}[H]
  \centering
  $\vcenter{\hbox{\includegraphics[scale=0.8]{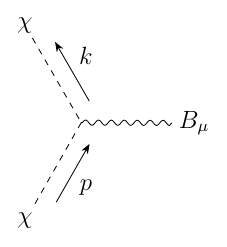}}}$
  $\displaystyle = - i \lambda_\chi (p + k)^\mu$.
\end{figure}

\noindent For the vector mediator the resulting scattering function is
\begin{align}
    \mathcal{K}(s, t) = \frac{1}{2}\Bigl[2m_p^2\left(m_k^2 - 2s + t\right) - 2 m_k^2 (2 s + t) + m_k^4 + m_p^4 + 4 s(s+t)\Bigr].
\end{align}
Analogously, for the axial vector mediator we obtain
\begin{align}
    \mathcal{K}(s, t) = 4\Bigl[-2m_k^2\left(m_p^2 + s\right) - 2 m_p^2 s + m_k^4 + m_p^4 + s(s+t)\Bigr].
\end{align}

\section{Vector DM} \label{app:vb-calc}
For vector DM and scalar mediator we take the vertex rule
\begin{figure}[H]
  \centering
  $\vcenter{\hbox{\includegraphics[scale=0.8]{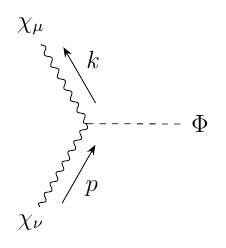}}}$
  $\displaystyle = - i m_\chi \lambda_\chi g^{\mu \nu}$.
\end{figure}

\noindent Again, we parametrize the mass scale appearing in the three boson vertex as $m_\chi$. In a UV-complete model, this convention can be recovered via redefinition of the coupling constant. We obtain the scattering function
\begin{align}
    \mathcal{K}(s, t) = \frac{2}{3 m_p^2} \left(2 m_k^2 - \frac t2\right) \left(\left(m_p^2 - \frac t2\right)^2 + 2 m_p^4\right).
\end{align}
Analogously, for a pseudoscalar mediator,
\begin{align}
    \mathcal{K}(s, t) = -\frac{ t}{3 m_p^2} \left(\left(m_p^2 - \frac t2\right)^2 + 2 m_p^4\right).
\end{align}

\noindent Finally, for the case of a vector DM and vector mediator we take the Feynman rule for the three vector vertex as

\begin{figure}[H]
  \centering
  $\vcenter{\hbox{\includegraphics[scale=0.8]{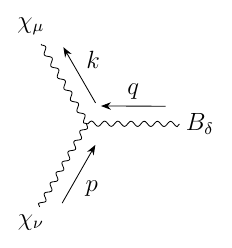}}}$
  $\displaystyle = - i \lambda_\chi \Bigl[-g^{\mu \nu}\left(p + k\right)^\delta + g^{\nu \delta}\left(p - q\right)^\mu + g^{\delta \mu}\left(q + k\right)^\nu\Bigr].  \nonumber$.
\end{figure}

\noindent The resulting scattering function is given by
\begin{align}
    \mathcal{K}(s, t) = &\frac{1}{3 m_{p }^4}\Bigl[m_{p }^4 \Bigl(12 m_k^2 \left(m_k^2-2 s+t\right)
    +12 s^2+20 s t+17 t^2\Bigr) 
-2 t m_{p }^2 \left(m_k^2 (t-4 s)+2 m_k^4 +2 s^2+3 s t+2 t^2\right) \nonumber \\
&+t^2 \left(s-m_k^2\right) \left(-m_k^2+s+t\right)
-4 m_{p }^6 \left(-6 m_k^2+6 s+t\right)+12 m_{p }^8\Bigr].
\end{align}
Analogously, for an axial vector mediator,
\begin{align}
        \mathcal{K}(s, t) = &\frac{1}{3 m_{p }^4}\Bigl[m_{p }^4 \Bigl(-8 m_k^2 \left(3 s+7t\right) +12 m_k^4 +12 s^2 + 20 st+17 t^2\Bigr) 
    -2 t m_{p }^2 \left(-m_k^2 (4 s + 7t)+2 m_k^4 +2 s^2+3 s t+2 t^2\right)\nonumber \\
    &+t^2 \left(-2 s m_k^2 + m_k^4 + s(s+t)\right)
    -4 m_{p }^6 \left(-6 m_k^2+6 s+t\right)+12 m_{p }^8\Bigr].
\end{align}

\end{document}

%% file: COSINUS.tex
\author{G.~Angloher}
\affiliation{Max-Planck-Institut f\"ur Physik, 85748 Garching - Germany}

\author{M.~R.~Bharadwaj}
\affiliation{Max-Planck-Institut f\"ur Physik, 85748 Garching - Germany}

\author{A.~B\"ohmer}
\affiliation{Marietta-Blau-Institut f\"ur Teilchenphysik der \"Osterreichischen Akademie der Wissenschaften, 1050 Wien - Austria}
\affiliation{Atominstitut, Technische Universit\"at Wien, 1020 Wien - Austria}

\author{M.~Cababie}
\affiliation{Marietta-Blau-Institut f\"ur Teilchenphysik der \"Osterreichischen Akademie der Wissenschaften, 1050 Wien - Austria}
\affiliation{Atominstitut, Technische Universit\"at Wien, 1020 Wien - Austria}

\author{I.~Colantoni}
\affiliation{INFN - Sezione di Roma, 00185 Roma - Italy}

\author{I.~Dafinei}
\affiliation{INFN - Sezione di Roma, 00185 Roma - Italy}
\affiliation{Gran Sasso Science Institute, 67100 L'Aquila - Italy}

\author{N.~Di~Marco}
\affiliation{Gran Sasso Science Institute, 67100 L'Aquila - Italy}
\affiliation{INFN - Laboratori Nazionali del Gran Sasso, 67100 Assergi - Italy}

\author{C.~Dittmar}
\affiliation{Max-Planck-Institut f\"ur Physik, 85748 Garching - Germany}

\author{F.~Ferella}
\affiliation{INFN - Laboratori Nazionali del Gran Sasso, 67100 Assergi - Italy}

\author{F.~Ferroni}
\affiliation{Gran Sasso Science Institute, 67100 L'Aquila - Italy}
\affiliation{INFN - Sezione di Roma, 00185 Roma - Italy}

\author{S.~Fichtinger}
\affiliation{Marietta-Blau-Institut f\"ur Teilchenphysik der \"Osterreichischen Akademie der Wissenschaften, 1050 Wien - Austria}

\author{A.~Filipponi}
\affiliation{INFN - Laboratori Nazionali del Gran Sasso, 67100 Assergi - Italy}
\affiliation{Dipartimento di Scienze Fisiche e Chimiche, Universit\`a degli Studi dell'Aquila, 67100 L'Aquila - Italy}

\author{T.~Frank}
\affiliation{Max-Planck-Institut f\"ur Physik, 85748 Garching - Germany}

\author{M.~Friedl}
\affiliation{Marietta-Blau-Institut f\"ur Teilchenphysik der \"Osterreichischen Akademie der Wissenschaften, 1050 Wien - Austria}

\author{D. Fuchs}
\affiliation{Marietta-Blau-Institut f\"ur Teilchenphysik der \"Osterreichischen Akademie der Wissenschaften, 1050 Wien - Austria}
\affiliation{Atominstitut, Technische Universit\"at Wien, 1020 Wien - Austria}

\author{L.~Gai}
\affiliation{State Key Laboratory of Functional Crystals and Devices, Shanghai Institute of Ceramics, Chinese Academy of Sciences, 201899 Shanghai, China}

\author{M.~Gapp}
\affiliation{Max-Planck-Institut f\"ur Physik, 85748 Garching - Germany}

\author{M.~Heikinheimo}
\email{matti.heikinheimo@helsinki.fi}
\affiliation{Helsinki Institute of Physics, 00014 University of Helsinki - Finland}

\author{M.~N.~Hughes}
\affiliation{Max-Planck-Institut f\"ur Physik, 85748 Garching - Germany}

\author{K.~Huitu}
\affiliation{Helsinki Institute of Physics, 00014 University of Helsinki - Finland}

\author{M.~Kellermann}
\affiliation{Marietta-Blau-Institut f\"ur Teilchenphysik der \"Osterreichischen Akademie der Wissenschaften, 1050 Wien - Austria}
\affiliation{Atominstitut, Technische Universit\"at Wien, 1020 Wien - Austria}

\author{R.~Maji}
\affiliation{Marietta-Blau-Institut f\"ur Teilchenphysik der \"Osterreichischen Akademie der Wissenschaften, 1050 Wien - Austria}
\affiliation{Atominstitut, Technische Universit\"at Wien, 1020 Wien - Austria}

\author{M.~Mancuso}
\affiliation{Max-Planck-Institut f\"ur Physik, 85748 Garching - Germany}

\author{L.~Pagnanini}
\affiliation{Gran Sasso Science Institute, 67100 L'Aquila - Italy}
\affiliation{INFN - Laboratori Nazionali del Gran Sasso, 67100 Assergi - Italy}

\author{F.~Petricca}
\affiliation{Max-Planck-Institut f\"ur Physik, 85748 Garching - Germany}

\author{S.~Pirro}
\affiliation{INFN - Laboratori Nazionali del Gran Sasso, 67100 Assergi - Italy}

\author{F.~Pr\"obst}
\affiliation{Max-Planck-Institut f\"ur Physik, 85748 Garching - Germany}

\author{G.~Profeta}
\affiliation{Dipartimento di Scienze Fisiche e Chimiche, Universit\`a degli Studi dell'Aquila, 67100 L'Aquila - Italy}
\affiliation{INFN - Laboratori Nazionali del Gran Sasso, 67100 Assergi - Italy}

\author{A.~Puiu}
\affiliation{INFN - Laboratori Nazionali del Gran Sasso, 67100 Assergi - Italy}

\author{F.~Reindl}
\affiliation{Marietta-Blau-Institut f\"ur Teilchenphysik der \"Osterreichischen Akademie der Wissenschaften, 1050 Wien - Austria}
\affiliation{Atominstitut, Technische Universit\"at Wien, 1020 Wien - Austria}

\author{K.~Sch\"affner}
\affiliation{Max-Planck-Institut f\"ur Physik, 85748 Garching - Germany}

\author{J.~Schieck}
\affiliation{Marietta-Blau-Institut f\"ur Teilchenphysik der \"Osterreichischen Akademie der Wissenschaften, 1050 Wien - Austria}
\affiliation{Atominstitut, Technische Universit\"at Wien, 1020 Wien - Austria}

\author{P.~Schreiner}
\affiliation{Marietta-Blau-Institut f\"ur Teilchenphysik der \"Osterreichischen Akademie der Wissenschaften, 1050 Wien - Austria}
\affiliation{Atominstitut, Technische Universit\"at Wien, 1020 Wien - Austria}

\author{C.~Schwertner}
\affiliation{Marietta-Blau-Institut f\"ur Teilchenphysik der \"Osterreichischen Akademie der Wissenschaften, 1050 Wien - Austria}
\affiliation{Atominstitut, Technische Universit\"at Wien, 1020 Wien - Austria}

\author{P.~Settembri}
\affiliation{Dipartimento di Scienze Fisiche e Chimiche, Universit\`a degli Studi dell'Aquila, 67100 L'Aquila - Italy}

\author{K.~Shera}
\affiliation{Max-Planck-Institut f\"ur Physik, 85748 Garching - Germany}

\author{M.~Stahlberg}
\affiliation{Max-Planck-Institut f\"ur Physik, 85748 Garching - Germany}

\author{A.~Stendahl}
\affiliation{Helsinki Institute of Physics, 00014 University of Helsinki - Finland}

\author{M.~Stukel}
\affiliation{SNOLAB, P3Y 1N2 Lively - Canada}
\affiliation{INFN - Laboratori Nazionali del Gran Sasso, 67100 Assergi - Italy}

\author{C.~Tresca}
\affiliation{CNR-SPIN c/o Dipartimento di Scienze Fisiche e Chimiche, Universit\`a degli Studi dell’Aquila, 67100 L’Aquila - Italy}
\affiliation{INFN - Laboratori Nazionali del Gran Sasso, 67100 Assergi - Italy}

\author{S.~Yue}
\affiliation{State Key Laboratory of Functional Crystals and Devices, Shanghai Institute of Ceramics, Chinese Academy of Sciences, 201899 Shanghai, China}

\author{V.~Zema}
\affiliation{Max-Planck-Institut f\"ur Physik, 85748 Garching - Germany}
\affiliation{Marietta-Blau-Institut f\"ur Teilchenphysik der \"Osterreichischen Akademie der Wissenschaften, 1050 Wien - Austria}

\author{Y.~Zhu}
\affiliation{State Key Laboratory of Functional Crystals and Devices, Shanghai Institute of Ceramics, Chinese Academy of Sciences, 201899 Shanghai, China}

\author{N.~Zimmermann}
\email{niklas.zimmermann@helsinki.fi}
\affiliation{Helsinki Institute of Physics, 00014 University of Helsinki - Finland}

\collaboration{The COSINUS Collaboration}\noaffiliation